\documentclass[12pt,letterpaper]{article}
\usepackage{bm}
\usepackage[dvips]{graphicx}
\usepackage{epic}
\usepackage{pst-plot}
\title{An Atom Trap Relying on Optical Pumping.
\thanks{{\em Europhys. Lett.}, {\bfseries 27}, (8), pp. 569-574, 10 September 1994.} }
\author{\textsc{P. Bouyer}, \textsc{P. Lemonde}, \textsc{M. Ben Dahan}, \textsc{A. Michaud}\\
\textsc{C. Salomon} and \textsc{J. Dalibard}\\
{\em Laboratoire Kastler Brossel, D\'{e}partement de Physique}\\
{\em de l'\'{e}cole Normale Sup\'{e}rieure and Coll\`{e}ge de France \footnote{Unit\'{e} de recherche de l'Ecole Normale Sup\'{e}rieure et de l'Universit\'{e} Pierre et Marie Curie, associ\'{e}e au CNRS (UA18).}}\\
{\em 24 rue Lhomond, 75231 Paris Cedex 05, France}\\}
\date{\normalsize (received 13 April 1994; accepted in final form 25 July 1994)}
\begin{document}
\maketitle


\begin{abstract}
We have investigated a new radiation pressure trap which relies on optical pumping and does not require any magnetic field. It employs six circularly polarized divergent beams and works on the red of a $J_{g} \longrightarrow J_{e} = J_{g} + 1$ atomic transition with $J_{g} \geq 1/2$. We have demonstrated this trap with cesium atoms from a vapour cell using the 852~nm $J_{g} = 4 \longrightarrow J_{e} = 5$ resonance transition. The trap contained up to $3\cdot 10^{7}$ atoms in a cloud of $1/\sqrt{e}$ radius of 330 $\mu$m.
\end{abstract}

The number of experiments using laser-cooled atoms has dramatically increased during the last few years, thanks to the relative easiness of capturing and cooling atoms with the radiation pressure force. The most commonly used device is the magneto-optical trap (MOT), which consists of 3 pairs of circularly polarized counterpropagating laser beams, superimposed on a gradient of magnetic field \cite{raab1987}. The Zeeman shifts for an atom located out of the centre of the trap cause an imbalance between the six radiation pressure forces, which results in a restoring force towards the centre of the trap.

\begin{figure}
\begin{center}
\includegraphics[scale=0.8]{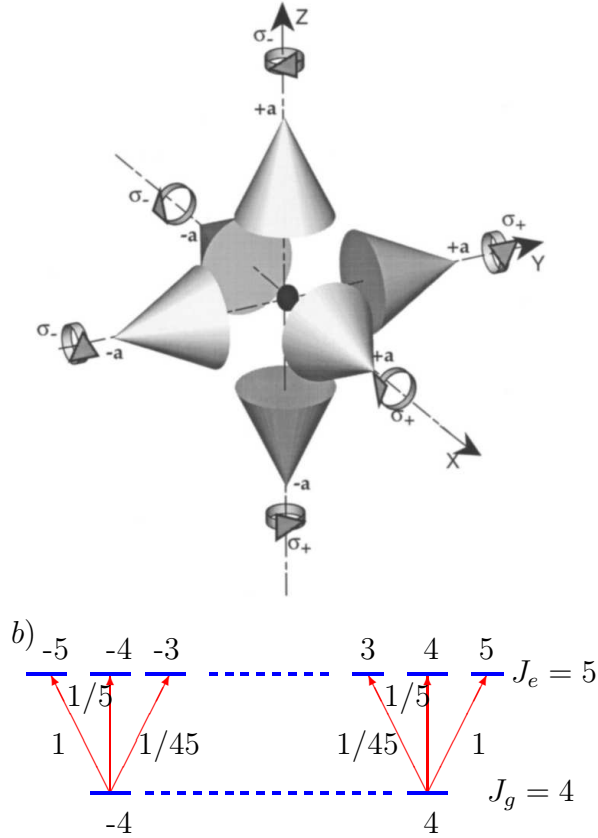}
\end{center}
\setlength{\unitlength}{3pt}
\begin{center}
\begin{picture}(70,25)
\put(0,24){$b)$}
\put(12.5,5){\red \vector(-1,2){7.5}}
\put(12.5,5){\red \vector(0,2){15}}
\put(12.5,5){\red \vector(1,2){7.5}}
\put(52.5,5){\red \vector(-1,2){7.5}}
\put(52.5,5){\red \vector(0,2){15}}
\put(52.5,5){\red \vector(1,2){7.5}}
\put(52,0){4}
\put(12,0){-4}
\put(4,22){-5}
\put(12,22){-4}
\put(18,22){-3}
\put(44,22){3}
\put(52,22){4}
\put(59,22){5}
\put(63,19){$J_{e}=5$}
\put(60,4){$J_{g}=4$}
\put(5,10){1}
\put(7,16){1/5}
\put(16,10){1/45}
\put(41,10){1/45}
\put(47,16){1/5}
\put(58,10){1}
\thicklines
\blue
\drawline(2,20)(7,20)
\drawline(10,20)(15,20)
\drawline(17,20)(22,20)
\drawline(25,20)(26,20)
\drawline(27,20)(28,20)
\drawline(29,20)(30,20)
\drawline(31,20)(32,20)
\drawline(33,20)(34,20)
\drawline(35,20)(36,20)
\drawline(37,20)(38,20)
\drawline(39,20)(40,20)
\drawline(43,20)(47,20)
\drawline(50,20)(55,20)
\drawline(58,20)(62,20)
\drawline(10,5)(15,5)
\drawline(17,5)(18,5)
\drawline(19,5)(20,5)
\drawline(21,5)(22,5)
\drawline(23,5)(24,5)
\drawline(25,5)(26,5)
\drawline(27,5)(28,5)
\drawline(29,5)(30,5)
\drawline(31,5)(32,5)
\drawline(33,5)(34,5)
\drawline(35,5)(36,5)
\drawline(37,5)(38,5)
\drawline(39,5)(40,5)
\drawline(41,5)(42,5)
\drawline(43,5)(44,5)
\drawline(45,5)(46,5)
\drawline(47,5)(48,5)
\drawline(50,5)(55,5)
\end{picture}
\end{center}
	\caption{Principle of the trap relying on optical pumping . $a$) Six circularly polarized divergent beams are detuned on the red of a  $J_{g} \longrightarrow J_{g}+1$ atomic transition with $J_{g} \ge 1/2$. Note the differences in helicity between the $Z$ beams and the $XY$ beams. The position-dependent optical pumping creates an imbalance in $m$ state populations. $b$) Since the squares  of the Clebsch-Gordan coefficients for $\sigma_{+}$ and $\sigma_{-}$ transitions are different, there is a global restoring force toward the centre. Shown here is the 4 $\longrightarrow$ 5 transition of cesium. In the experiment, each laser beam is focussed at 3.5 cm from the centre of the trap using an objective with NA=0.4.}
\label{figure:principle}
\end{figure}

We present in this letter a radiation pressure Trap Relying On Optical Pumping (TROOP), which does not require any magnetic field. Therefore it is particularly attractive for several applications such as cold-atom frequency standards \cite{kasevich1989}\cite{clairon1991} or subrecoil cooling \cite{aspect1988}\cite{kasevich1992}, where any residual magnetic field may cause severe limitations. The trap consists of six circularly polarized diverging laser beams (see fig. \ref{figure:principle}$a$)) which induce a position-dependent optical pumping resulting in a restoring force. The atomic transition used for the trapping process involves a ground state with angular momentum $J_{g}$ and an excited state with  $J_{e} = J_{g} + 1$. The trap shoud work for any $J_{g} \neq 0$. We have demonstrated this trapping effect for cesium atoms, using the  $J_{g} = 4 \longrightarrow J_{e} = 5$ transitions at 852 nm. The performance of the trap, in terms of capture of atoms from a vapour and confinement, are whitin one order of magnitude of the performance of the optimized MOT constructed with the same laser beams.

Soon after early proposals \cite{minogin1982}, the optical Earnshaw theorem was proved by Ashkin and Gordon, stating that no stable trapping could be achieved with the radiation pressure force for particles with \emph{scalar polarizabilities} \cite{ashkin1983}, such as an atom with a $J_{g} = 0 \longrightarrow J_{e} = 1$ transition. In weak, non-saturating laser light, the force is then proportional to the Poynting vector $\bm{\Pi}$. Since  $\bm{\nabla \cdot \Pi} = 0$, the force also has zero divergence. Therefore it cannot be an inward force everywhere on a closed surface.

A simple way to get round the Earnshaw theorem is to achieve a situation where the polarizability of the atom is not scalar. As first shown in \cite{pritchard1986}, a static magnetic field or a suitable optical-pumping configuration can lead to such non-scalar polarizabilities, and can, therefore, generate a stable radiation pressure trap. The MOT, although different from the explicit proposals of \cite{pritchard1986}, works along the same principle \cite{raab1987}.

The trap that we have investigated (fig. \ref{figure:principle}$a$)) also relies on a non-linear relation between the radiation pressure force and the Poynting vector, induced by optical pumping. We restrict our analysis to the low saturation limit: from the state of polarization of the total light field at a given point, we first derive the atomic internal steady state, where most of the population is in the ground state. Then we add independently the six radiation pressure forces acting on this atom. Our simple model neglects the interference effects between the beams, which may result in dipole or vortex forces on the wavelength scale.

Take an atom at point $A$, which is displaced a distance $z$ from the origin $O$ along the positive $Z$-axis. We assume that $z$ is small compared to the distance $a$ between $O$ and each focus of the diverging waves. The beam $W_{z+}$, propagating from $z=+\infty$ to $z=-\infty$, creates a radiation pressure force which tends to restore the atom towards $O$. The five other beams, \emph{i.e.} $W_{z-}$ and the four diverging waves propagating along the $X$ and $Y$ axes, produce a net expelling force along the $Z$-direction. For a particle with scalar polarizability, the restoring and expelling forces cancel. This is a manifestation of the Earnshaw theorem.

For circularly polarized laser waves with a convenient choice of helicities, and for an atomic transition having $J_{g} > 0 $, we now show that this theorem does not hold anymore if we take into account the modification of the internal state due to optical pumping. 

We restrict here to the case where the helicity $h_{i}$ of two waves $W_{i+}$ and $W_{i-}$ is the same, so that each pair of waves is in a $\sigma_{+}$ -- $\sigma_{-}$ configuration. At $O$, the total field has no prefered polarization axis and the ground-state population in the steady-state regime is equally distributed among the Zeeman sublevels.

Take the configuration where the $W_{z+}$ and $W_{z-}$ waves are respectively, $\sigma _{-}$ and $\sigma _{+}$ polarized along the the $Z$-axis; assume also that the polarizations of the waves propagating along $X$ and $Y$ have been chosen such that an atom located at $A$ sees a resulting light from the six beams with with a total $\sigma_{-}$ intensity $I(\sigma_{-})$ larger than then the $\sigma_{+}$ one. Because of optical pumping, the population of the ground-state sublevels $\left|g,m_{g}\right>$ with $m_{g} < 0$ increase and the populations for $m_{g}>0$ decrease. This breaks the balance between the six radiation pressure forces since, for $m_{g}<0$, the Clebsch-Gordan coefficient of the $\sigma_{-}$ transition starting from a given $\left|g,m_{g}\right>$ induced by the $W_{z+}$ wave is larger than the one for the $\sigma_{+}$ transition starting from the same substate and induced by the $W_{z-}$ wave (fig. \ref{figure:principle}$b$)). Therefore, to first order in $z/a$, one finds that the restoring force due to $W_{z+}$ is enhanced with respect to the case where no modification of optical pumping occurs; on the opposite, the expelling force due to $W_{z-}$ is reduced. Finally, the component on the $Z$-axis of the force due to the four $X$ and $Y$
beams which involves a geometrical $z/a$ factor remains unchanged to first order in $z/a$. The overall effect is a net restoring force towards $O$ along the $Z$-axis.

The previous discussion leads to a 3D trapping effect if one of the three helicities is different from the other two. If the three helicities are the same, then at first order in $z/a$ the light remains natural ($I(\sigma_{+})=I(\sigma_{-})=I(\pi)$). No trapping force can occur in this case since the steady state of the atom is unchanged at first order in $z/a$.

We choose for instance, as in fig \ref{figure:principle}$a$), $h_{z} = -h_{x} = -h_{y} = 1$; for an atom displaced a distance $z$ along the $Z$-axis, and for a quantization axis along $Z$, we obtain the relation between the intensities of the $\sigma_{\pm}$ and $\pi$ components of the lignt:

\begin{equation}
I(\sigma_{+})=\sum\limits_{X,Y,Z}\ I(\sigma_{+})= (1-2z/a)I(\pi),\ I(\sigma_{-})= (1+2z/a)I(\pi). 
\label{equation:intensityZ}
\end{equation}  
 
For an atom displaced by a quantity $\xi$ along the $X$ or $Y$ axis, we choose now the quantization axis along the displacement axis and we obtain
 
\begin{equation}
I(\sigma_{+})=\ (1+\xi / a )I(\pi),\ I(\sigma_{-})= (1- \xi /a)I(\pi) \ . 
\label{equation:intensitiesXY}
\end{equation}  

Consequently, this configuration, analogous to the one used in the MOT, guarantees that the light at a location different from the centre of the trap has a deviation from natural light which favours the restoring component of the force. We can write the force for an atom, respectively, located on the $Z$ or on the $XY$-axis:

\begin{equation}
f_{z}\ =\  - 2 \mu \  \frac{z}{a} \  F_{0}\ ,\ \ \ \ f_{\xi}\  =\  - \mu \  \frac{\xi }{a} \ F_{0}\ .
\label{equation:forcecomponent}
\end{equation}  

$F_{0}$ is the force exerted by a single wave on an atom at rest in $O$:

\begin{equation}
F_{0}\  =\  \hbar k \Gamma\  \frac{\Omega^{2}}{\Gamma^{2}+4\delta^{2}}\ ,
\label{equation:singleforce}
\end{equation}  

where $k$ is the wave vector of the light, $\gamma$ is the natural width of the atomic excited state, $\Omega$ is the Rabi frequency at $O$ calculated for a transition with a Clebsch-Gordan coefficient equal to 1, and $\delta=\omega_{L}-\omega_{A}$ is the detuning between the laser and atomic frequencies. Since $f_{z}=2f_{\xi}$, we expect the trap to be elliptic, the ratio between major and minor axis being $\sqrt{2}$. In (3), the numerical coefficient $\mu$ which is a fraction of unity, characterizes the efficiency of the trapping process and at small saturation it is expected to be independent of intensity. A quantitative evaluation of $\mu$ and its dependence on the laser parameters would require a complete determination of the atomic steady state in the 3D optical field. One would then recover the neglected dipole force contributions as well as corrections due to the spatial modulation on the wavelength scale of the optical-pumping processes.

We have demonstrated this trap in an apparatus similar to that used for the usual vapour cell MOT \cite{monroe1990}. Three laser diodes generate the three beam pairs along $X$,$Y$,$Z$. They are injected by a grating-tuned extended-cavity laser diode. The total power in each beam pair is tuned using a half-wave plate polarizing cube system before being split into two counterpropagating beams using a second half-wave plate polarizing cube system. This allows a fine tuning of the intensity balance among the six trap beams. Each beam is focused 3.5 cm from the centre of the trap and has a $\pm 22^{\circ}$ divergence. The polarization is set circular with an ellipticity $(I(\sigma_{+})-I(\sigma_{-})/I(\sigma_{+})+I(\sigma_{-}))$ less than 4\%. The Earth's magnetic field is compensated to less than 10 mG. Optional MOT coils are installed, since a convenient way for finding the TROOP is to start from a MOT and gradually decrease the magnetic-field gradient, while balancing precisely the six laser beam intensities. The coils also allow an easy comparison between the TROOP and the MOT.

We have measured the number of trapped atoms and the cloud dimensions with a CCD camera looking from the (1,1,1) direction. The atom number is deduced (to whitin a factor of $\sim 2$ ) from the fluorescence emited by the trapped atoms. A time-of-flight method gives access to the vertical temperature, using a probe beam located 5 cm below the trap. In order to let the atoms fall, the current of the injected diodes is quickly ($\leq$ 100 ns) changed so as to destroy the injection locking. This switches their wavelength far away from the atomic resonance. In a second stage, the diodes are turned off in about 50 $\mu$s.

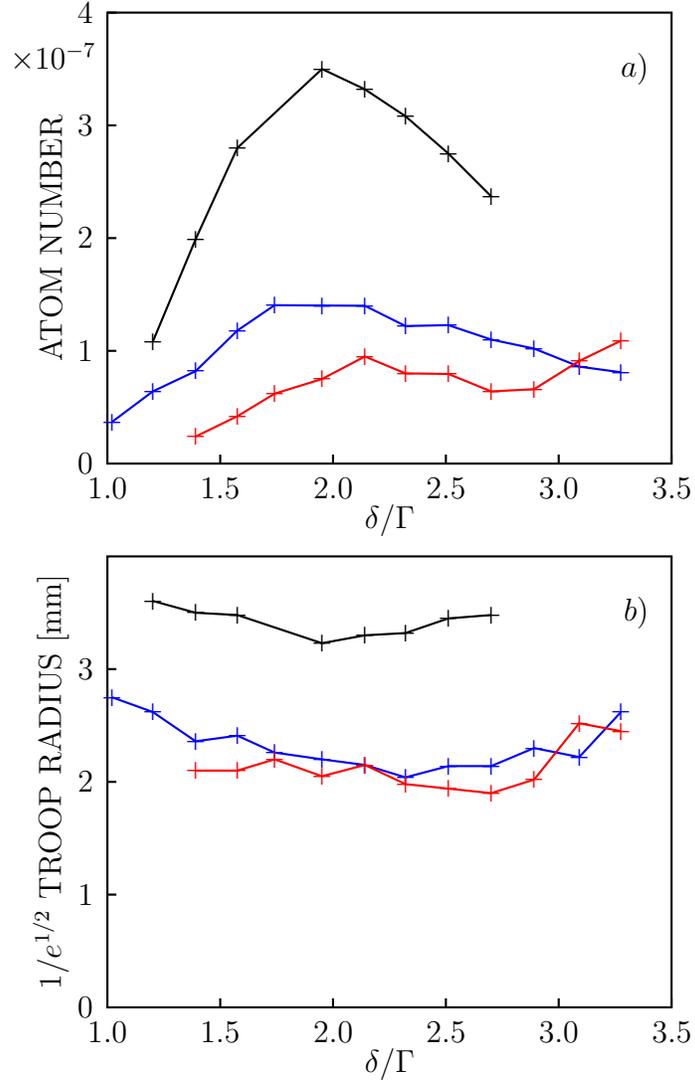
\begin{figure}
\begin{center}
\psset{xunit=3cm,yunit=1.5cm}
\begin{pspicture}(0.7,-0.8)(3.5,4)
\psaxes[Ox=1.0,Dx=0.5,Oy=0,Dy=1,axesstyle=frame,tickstyle=top](1.0,0)(3.5,4)
\savedata{\mydata}[{
{1.2,1.08},
{1.39,1.99},
{1.575,2.8},
{1.95,3.5},
{2.14,3.32},
{2.32,3.081},
{2.51,2.75},
{2.7,2.37}
}]
\dataplot[plotstyle=line,showpoints=true,dotstyle=+,dotscale=1.5.0,linecolor=black]{\mydata}
\savedata{\mydata}[{
{1.02,0.37}
{1.2,0.64},
{1.39,0.82},
{1.575,1.18},
{1.74,1.405}
{1.95,1.402},
{2.14,1.4},
{2.32,1.22},
{2.51,1.228},
{2.7,1.1},
{2.89,1.02},
{3.09,0.86},
{3.275,0.81}
}]
\dataplot[plotstyle=line,showpoints=true,dotstyle=+,dotscale=1.5,linecolor=blue]{\mydata}
\savedata{\mydata}[{
{1.39,0.24},
{1.575,0.42},
{1.74,0.62}
{1.95,0.75},
{2.14,0.95},
{2.32,0.8},
{2.51,0.795},
{2.7,0.64},
{2.89,0.66},
{3.09,0.91},
{3.275,1.09}
}]
\dataplot[plotstyle=line,showpoints=true,dotstyle=+,dotscale=1.5,linecolor=red]{\mydata}
\rput*[r](3.4,3.5){$a)$}
\rput*[c](2.25,-0.5){$\delta / \Gamma $}
\rput*[c]{L}(0.75,2.0){ATOM NUMBER}
\rput*[r](0.95,3.6){$\times 10^{-7}$}
\end{pspicture}
%
\psset{xunit=3cm,yunit=15cm}
\begin{pspicture}(0.7,-0.08)(3.5,0.4)
\psaxes[Ox=1.0,Dx=0.5,Oy=0,Dy=0.1,axesstyle=frame,tickstyle=top](1.0,0)(3.5,0.4)
\savedata{\mydata}[{
{1.2,0.3605},
{1.39,0.35},
{1.575,0.348},
{1.95,0.323},
{2.14,0.33},
{2.32,0.332},
{2.51,0.345},
{2.7,0.348}
}]
\dataplot[plotstyle=line,showpoints=true,dotstyle=+,dotscale=1.5.0,linecolor=black]{\mydata}
\savedata{\mydata}[{
{1.02,0.275}
{1.2,0.262},
{1.39,0.236},
{1.575,0.241},
{1.74,0.226}
{1.95,0.22},
{2.14,0.215},
{2.32,0.204},
{2.51,0.214},
{2.7,0.214},
{2.89,0.23},
{3.09,0.222},
{3.275,0.262}
}]
\dataplot[plotstyle=line,showpoints=true,dotstyle=+,dotscale=1.5,linecolor=blue]{\mydata}
\savedata{\mydata}[{
{1.39,0.21},
{1.575,0.21},
{1.74,0.22}
{1.95,0.205},
{2.14,0.215},
{2.32,0.198},
{2.51,0.194},
{2.7,0.190},
{2.89,0.202},
{3.09,0.252},
{3.275,0.245}
}]
\dataplot[plotstyle=line,showpoints=true,dotstyle=+,dotscale=1.5,linecolor=red]{\mydata}
\rput*[r](3.4,0.35){$b)$}
\rput*[c](2.25,-0.05){$\delta / \Gamma $}
\rput*[c]{L}(0.75,0.2){$1/e^{1/2}$ TROOP RADIUS [mm]}
\end{pspicture}
\caption{Experimental results: the trap parameters ($a$) atom number and $b$) $1/ \sqrt{e}$ minor axis of the ellipse) are measured with a CCD camera looking from the (1, 1, 1)-direction, for 3 different single wave Rabi frequencies 
$\Omega=0.8\cdot \Gamma$~({\black \large +}), $\Omega= \Gamma$~({\blue \large +}), $\Omega=1.5\cdot \Gamma$~({\red \large +}). The peak density in the trap is $\sim 3 \cdot 10^{10}$ at. cm$^{-3}$.}   
\label{figure:results}
\end{center}
\end{figure}

As shown in fig. \ref{figure:results}, at a detuning of $-2\Gamma$, a Rabi frequency of 0.8 ($\pm$ 0.12)$\Gamma$ per wave and a vapour pressure of $\sim 5\cdot 10^{-8}$ Torr, our trap contained up to $3 \cdot 10^{7}$ atoms. There is a pronounced optimum for the trapped-atom number at $|\delta| \sim 2\Gamma$, whereas the trap size does not vary much with the detuning for $\Gamma \leq |\delta| \leq 3\Gamma$. The trapped-atoms cloud is found to be of elliptic shape, the minor axis being along the $Z$-direction. The position distribution along the minor axis of the CCD camera picture was well fitted by a Gaussian curve having a $1/\sqrt{e}$ radius of 330 ($\pm$ 50) $\mu$m. By analysing several trap pictures for different laser parameters, we find that the ratio between major and minor axis is 1.3 ($\pm$0.1). Taking into account the direction of observation, this leads to a ratio of $\sim$ 1.5 ($\pm$ 0.1) between the trap size along $XY$-axis and along the $Z$-axis. This is in good agreement with the $\sqrt{2}$ value expected from (\ref{equation:forcecomponent}) and (\ref{equation:singleforce}). The peak atomic density is $\sim 3 \cdot 10^{10}$ at. cm$^{-3}$. At these densities, we did not notice any significant deviation from a spatial Gaussian distribution, that one should expect if atom-atom interactions played an important role \cite{walker1990}\cite{steane1991}\cite{drewsen1994}.
%

We made several additional tests to confirm our understanding of this new trapping mechanism. First, the trap disappears if all six waves have the same helicity: when we invert the two helicities of the beams along the $Z$-axis, the trap vanishes, leaving only a uniform molasses signal. From this situation, if we now invert the helicities of the other beams along the $X$ and $Y$ axes, we recover the TROOP. This excludes trapping through any resisual magnetic-field gradient. Secondly, the intensity balance in each trapping beam pair is very critical and must be done at the percent level. This trap is also sensitive to the intensity imbalance between the $X$, $Y$ and $Z$ pairs. A 20\% intensity increase in the $X$ beams decreases the number of trapped atomss by 25\%, as the expelling effect of these beams along the $Y$ and $Z$ axes gradually compensates the restoring force.

In order to estimate the trap spring constant, $\kappa$, we take the data at small saturation $(\Omega = 0.8 \ \Gamma,\ \delta=-2 \ \Gamma)$. From the major axis of 400 ($\pm$50) $\mu$m,  from a temperature of 40 ($\pm$10) $\mu$K, and from
\begin{equation}
\kappa \  = \ \frac{k_{B}T}{r^{2}}\ ,
\end{equation}  
we find $\kappa \sim 5 \cdot 10^{-21}$ J/m$^{2}$, a value which is one order of machitude below the sping constant of a MOT at the same intensity and detuning and a gradient of 10 G/cm along the coil axis \cite{steane1991},\cite{drewsen1994}. From
\begin{equation}
\mu \ = \ \frac{a \kappa}{F_{0}}\ ,
\end{equation}  
we find $\mu \sim 0.1$.

Finally we found that the number of atoms in the TROOP was 5 to 20 times lower than in an optimized vapour cell MOT at the same detuning. Rabi frequency and confinement volume. Indeed, the Zeeman assisted slowing effect occurs during the capture process of a MOT and enhances the maximum velocity which can be effectively cooled. For loading a larger number of atoms in the TROOP, the use of pre-cooled atoms such as slow atomic beam using the frequency-chirping method or atoms falling from a spacially separated MOT may be a better choice. In the cell trap, one could also think of frequency-chirping the trapping beams and/or using additional laser beams. Finally the role of the relative phases between the beams on this trapping mechanism should be further investigated theoritically theoritically and experimentally using either servo-controled phases \cite{hemmerich1992} or the minimal four-beam geometry of \cite{grynberg1993}.

To summarize, we believe that the simplicity of this trap and the absence of magnetic field will make it particularly usefull in various applications.

\begin{center}
*\quad *\quad *
\end{center} 

We wish to acknowledge usefull discussions with \textsc{Y. Castin}, \textsc{A. Steane}, and our colleagues of the ENS laser cooling group. This work was supported in part by BNM, CNES, DRET, Coll\`{e}ge de France, and NEDO (Japan).

\end{document}